# Effect of stoichiometry on oxygen incorporation in MgB$_2$ thin films

**R K Singh, Y Shen, R Gandikota, J M Rowell and N Newman**
School of Materials, Arizona State University, Tempe, Arizona 85287-6006, USA

E-mail: Nathan.Newman@asu.edu

**Abstract.** The amount of oxygen incorporated into MgB$_2$ thin films upon exposure to atmospheric gasses is found to depend strongly on the material's stoichiometry. Rutherford backscattering spectrometry was used to monitor changes in oxygen incorporation resulting from exposure to: (a) ambient atmosphere, (b) humid atmospheres, (c) anneals in air and (d) anneals in oxygen. The study investigated thin-film samples with compositions that were systematically varied from Mg$_{0.9}$B$_2$ to Mg$_{1.1}$B$_2$. A significant surface oxygen contamination was observed in all of these films. The oxygen content in the bulk of the film, on the other hand, increased significantly only in Mg rich films and in films exposed to humid atmospheres.

## 1. Introduction

Magnesium diboride (MgB$_2$), with a high transition temperature (40 K) [1] and upper critical field (>50 T) in carbon doped films [2], has the potential to replace low transition temperature (T$_c$) materials (NbTi, Nb$_3$Sn) in high-field applications. One issue in the transport properties of superconductive MgB$_2$ is its apparent sample-dependency. It has been pointed out that reduced connectivity in MgB$_2$ samples (both films and bulk polycrystalline) results in extremely large variations in resistivity values [3]. If connectivity is small, only a fraction of the sample carries current, or supercurrent. Hence the value of apparent resistivity at all temperatures is increased over the values observed in single crystals, and also in the cleanest bulk [4] and thin film samples [5]. In particular the difference in resistivity from 300 K ($\rho_{300}$) to the value measured at 40 K ($\rho_{40}$), i.e. $\Delta\rho = \rho_{300} - \rho_{40}$, when compared to the value in clean films, has been used as a measure of sample connectivity. Within this simple picture, the residual resistivity ratio (RRR) ($\rho_{300}/\rho_{40}$) is not changed by variations in connectivity, but is decreased when increased defect scattering increases $\rho_{40}$.

MgB$_2$ films readily pick up oxygen from the ambient atmosphere forming various oxides (BO$_x$, MgO, MgBO, BO$_x$-MgO$_y$ etc.) [6,7]. It has been suspected that oxygen in MgB$_2$ can affect the resistivity in two primary ways. If it forms MgO in the grain boundaries, connectivity will be decreased and $\Delta\rho$ increased. If small particles of MgO act as scattering centers within the grains themselves, then $\rho_{40}$ will be increased, connectivity will be unchanged and RRR will decrease. The reported resistivity of thin films at room temperature, $\rho(300\text{ K})$, ranges from less than 10 to more than 100 $\mu\Omega.\text{cm}$, and that at 40 K, $\rho(40\text{ K})$, from less than 1 to more than 80 $\mu\Omega.\text{cm}$ [8]. Much higher values have been reported in other $MgB_2$ studies [3]. While resistivity appears to depend on grain coupling (connectivity) in many cases, the large variations in RRR suggest that dense lattice defects may exist within $MgB_2$ grains. Zhu et al. [9] have shown that MgO can exist inside $MgB_2$ grains as stacking faults and create microstrains. These oxygen-enriched phases can act as pinning centers, improving the critical current density of $MgB_2$ film.

We have observed significant variation in the transition temperature ($T_c$), resistivity ($\rho$), and upper critical field ($H_{c2}$) as a function of oxygen concentration in films [10]. A systematic correlation among residual resistivity ratio (RRR), the lattice distortion (microstrain) and oxygen-concentration reported by Xue et al. [11] suggests that the transport properties of $MgB_2$ may indeed be dominated by oxygen-related defects. Oxygen contamination also becomes critical as $MgB_2$ films have an inherent tendency to absorb oxygen from ambient atmosphere during storage. This affects their superconducting properties and temporal stability.

The intention of this work is to clarify the mechanism of oxygen incorporation in thin films and its role in determining the superconducting properties. In this study we report the effect of film stoichiometry and ambient conditions on oxygen incorporation in $MgB_2$ films.

## 2. Experiment

1500 – 2000 Å thick $MgB_2$ films were deposited on (0001) sapphire at 300 ± 2 $^o$C in an ultra-high vacuum MBE system with an ultimate base pressure of $\sim 5 \times 10^{-10}$ Torr. The system pressure reached as high as $10^{-6}$ Torr during deposition. Further growth details have been published elsewhere [12, 13]. The composition of the films was varied from $Mg_{0.9}B_2$ to $Mg_{1.1}B_2$ by controlling the Mg flux during growth. "Standard" films were cooled to room temperature in the chamber and then vented with nitrogen to atmospheric pressure. For "oxygen annealed" films, the growth chamber was vented with oxygen (200 mTorr to 1 atmosphere) at 250 $^o$C. "Air annealed" films were made by heating the "standard" films in ambient atmosphere to 300 $^o$C. For comparison, we also studied stoichiometric films grown at 500 $^o$C and 700 $^o$C at Superconductor Technologies Inc. (STI) and Pennsylvania State University (PSU) respectively.

The thicknesses and depth profiles of the chemical composition in all the films were determined using Rutherford backscattering spectrometry (RBS). Nuclear resonant elastic scattering, $^{16}O(\alpha,\alpha)^{16}O$ was used for oxygen detection in (a) standard, (b) oxygen annealed, and (c) air annealed films. RBS can detect up to 0.5% of oxygen with an accuracy of >95%. This technique, however, does not indicate whether impurity elements are in the grains or in the grain boundaries. Typical RBS data and analysis of an $MgB_2$ film with 0.5% oxygen content in the bulk and 5% oxygen content at the surface is shown in figure 1.

Studies were also made to quantify the extent of oxygen incorporation in $Mg_xB_2$ films stored in (a) ambient and (b) humid atmospheres. For these studies, some films were kept in an ambient atmosphere for several months and others were kept in a humid atmosphere (relative humidity (RH) of 50-70%) for several weeks.

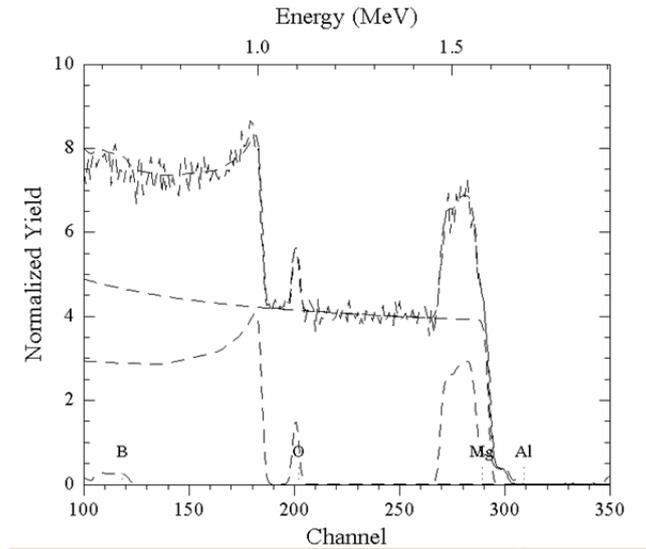

**Figure 1.** RBS data (solid) and simulation (dashed) of an $MgB_2$ thin film showing 0.5% oxygen in bulk and 5% oxygen in top 90 Å layer.

### 3. Results and discussion

*3.1 Effect of growth conditions*

The extent of degradation of films produced in various labs has been found to vary markedly. We have routinely studied films made at Arizona State University (ASU), Pennsylvania State University (PSU), and Superconductor Technologies, Inc. (STI). We have found that films produced at STI with a reactive evaporation process can be stored under atmospheric conditions without significant changes in their electrical and chemical properties for months and possibly even longer. X-ray photoelectron spectroscopy (XPS) studies of such STI films grown at 500 °C show a thin layer (20-30 Å) of Mg and B oxides on the surface [14]. The XPS measurements made several months after the films were deposited indicated that the surface layer remained thin and was not degraded [14]. In contrast, we found that the highest-quality (with lowest $\rho_{40K}$ values) films grown at 700 °C at PSU undergo significant changes in the chemical and electrical properties within a week. After a few weeks, the material becomes transparent and does not exhibit superconductivity. The films grown at ASU at 300 °C using MBE fall somewhere between these extremes, degrading slowly with time. It appears, therefore, that the stability of $MgB_2$ films greatly depends upon the growth mechanism, perhaps on the surface stoichiometry and not necessarily on the substrate temperature during growth. The mechanism for this wide range of behavior has not been fully elucidated.

*3.2 MgB$_2$ films grown in MBE at ASU*

A summary of oxygen and carbon content in MgB$_2$ films before and after various treatments is presented in table 1. Oxygen contamination in as-grown films is strongly dependent on the film stoichiometry. "Surface" oxygen in Mg-deficient (Mg$_{0.9}$B$_2$), stoichiometric (MgB$_2$), and Mg-rich (Mg$_{1.1}$B$_2$) film is typically 2%, 3% and 12% respectively. As shown in figure 2, there is considerable scatter in contamination levels. The thickness of the contaminated surface layer varies from 30 to 100 Å. Oxygen in the bulk of the film (as-grown), on the other hand, is typically less than 1% except for Mg-rich films where bulk oxygen as high as 10% has been observed, see figure 2. This indicates that the surface oxygen contamination occurred on exposure to air and/or during post-growth storage. The low solid solubility of oxygen in MgB$_2$ at room temperature rules out the possibility of atmospheric oxygen diffusing into the MgB$_2$ lattice [15]. The moisture present in the atmosphere might be expected to form (Mg,B) hydroxides at the surface. The presence of a significant amount of carbon (1 – 3 %) at the surface suggests that these hydroxides might have then further reacted with the CO$_2$ present in the atmosphere, leaving carbon residue on the surface. We could not detect any near-surface phases by X-ray diffraction (XRD) studies as they are present only in a very thin layer on the film.

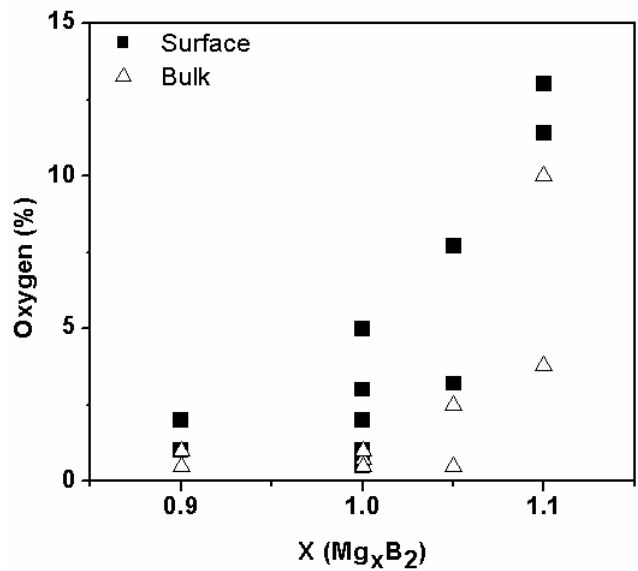

**Figure 2.** Effect of film stoichiometry on oxygen incorporation in as-grown films.

Table 1. Surface and bulk contamination of stoichiometric MgB$_2$ after various treatments.

| # | Treatment | Oxygen Surface | Oxygen Bulk | Carbon Surface | Carbon Bulk | Remarks |
|---|---|---|---|---|---|---|
| 1 | As grown (300 °C in MBE) | 5% in top 100 Å | 0.5% | 1.5% in top 100 Å | Not detected | - |
| 2 | Oxygen annealed (10.5 Torr/250 °C) | 12% in top 300 Å | 2% | 1.5% in top 100 Å | Not detected | - |
| 3 | Air annealed (200 °C) | 15% in top 300 Å | 6% | 5% in top 200 Å | 1% | - |
| 4 | Exposed to (acetone + photoresist) vapour for 20 hours | 3% in top 200 Å | 0.5% | 12.5% in top 100 Å | Not detected | - |
| 5 | Dipped in acetone for 60 hours | 3% in top 200 Å | 0.5% | 20% in top 200 Å | Not detected | - |
| 6 | Exposed to water vapour for 68 hours | Completely degraded to oxides, hydrides and hydroxides of Mg and B | | | | No superconductivity observed |
| 7 | Dipped in (water + acetone) for 18 hours | 6% in top 450 Å | 1% | 9% in top 450 Å | Not detected | Etch rate ~ 10 Å/min (ASU), ~ 1 Å/min (STI) |

*3.3 Effect of storage and post-growth treatments*

*3.3.1 Effect of storage in ambient atmosphere.* Surface contamination increases with storage time and is dependent on the film stoichiometry. After 6 months of storage of ASU films, a contaminated layer thickness of ~100 Å, ~200 Å, and ~300 Å was observed for Mg-deficient, stoichiometric, and Mg-rich films, respectively. This indicates that Mg has a higher affinity for atmospheric oxygen than boron does, as might be anticipated. Optical studies of these samples indicate that the degradation takes place preferentially at the grain boundaries (figure 3). This makes fine grained MgB$_2$ more susceptible to atmospheric degradation. Since all the ASU films in this study were grown under identical conditions and had similar grain size, degradation was mainly controlled by the stoichiometry of the films. Appearance of a dendritic pattern after degradation, as shown in figure 4, suggests microsegregation of Mg along dendritic walls during film growth.

*3.3.2 Effect of post growth annealing and humidity exposure.* In an attempt to understand the kinetics of oxygen contamination, MgB$_2$ films were subjected to various adverse conditions. In humid atmospheres (50-70% RH), these ASU films completely oxidize in less than 3 days and no longer exhibit a superconducting transition. These films, however, showed a strong dependence on composition when annealed in air or oxygen. As shown in figure 5, Mg-rich films annealed in 10.5 Torr at 250 $^o$C show surface oxygen of > 25% and bulk oxygen as high as 20%. On the other hand, bulk oxygen of <3% was observed in stoichiometric films after a similar annealing treatment.

A similar trend was observed after air-annealing $MgB_2$ films at 200 $^o$C. Air annealed Mg-rich samples also had up to 3% carbon along with oxygen in the bulk film (figure 6). Mg-rich films, after annealing, have fairly uniform oxygen distribution with thickness. Stoichiometric and Mg-deficient films, on the other hand, show a layered structure after air annealing. The top half of the film is found to have comparable oxygen content to Mg-rich films, although the content of the lower half of the film is not significantly affected by annealing.

These results imply need for passivation of $MgB_2$ films. Protective overcoats may provide long term stability during storage, processing and device operation. We have used protective overcoats of $Ta_xN$, $SiO_2$ and Au on $MgB_2$ films. Although all these protective overcoats are effective in minimizing the film degradation, $Ta_xN$ in particular looks attractive as it has a resistivity (1.9 mΩ.cm at room temperature and 11 mΩ.cm at 4.2 K) [16] high enough to not significantly influence the electrical measurements while still allowing contacts to be made easily.

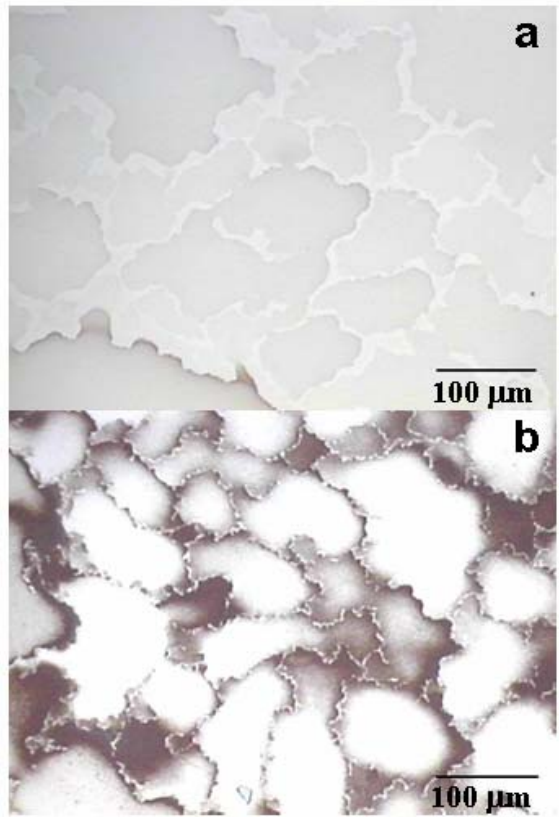

**Figure 3.** Degradation of $MgB_2$ film in ambient atmosphere after (a) 1 week and (b) 1 month.

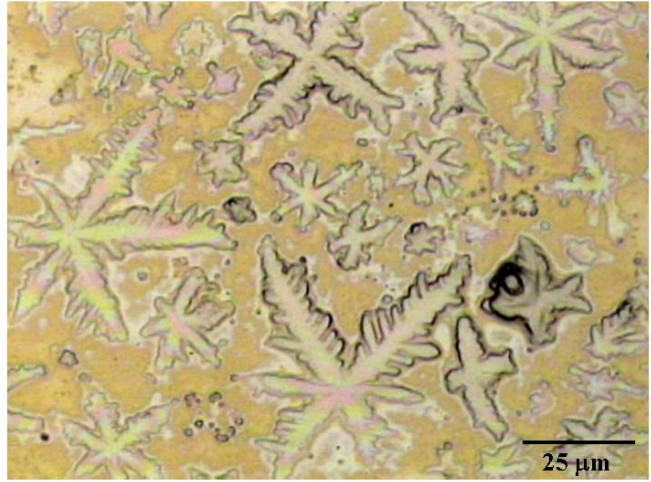

**Figure 4.** Dendritic pattern at the surface of degraded MgB$_2$. Note the four-fold and six-fold symmetry.

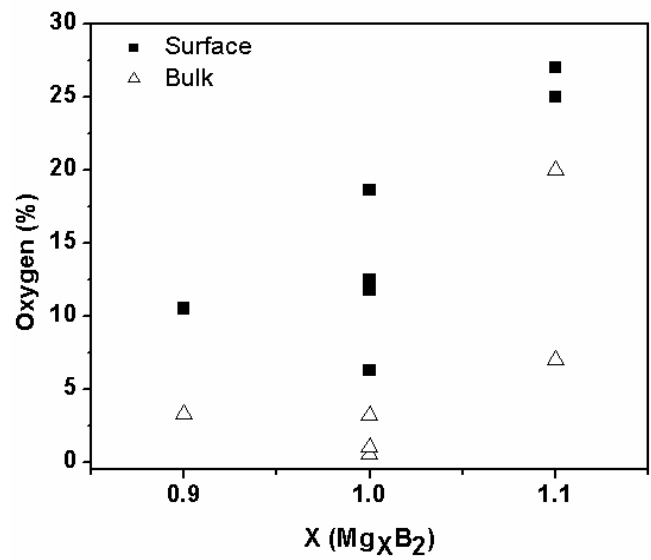

**Figure 5.** Effect of film stoichiometry on oxygen incorporation after annealing in 10.5 Torr of oxygen at 250 °C.

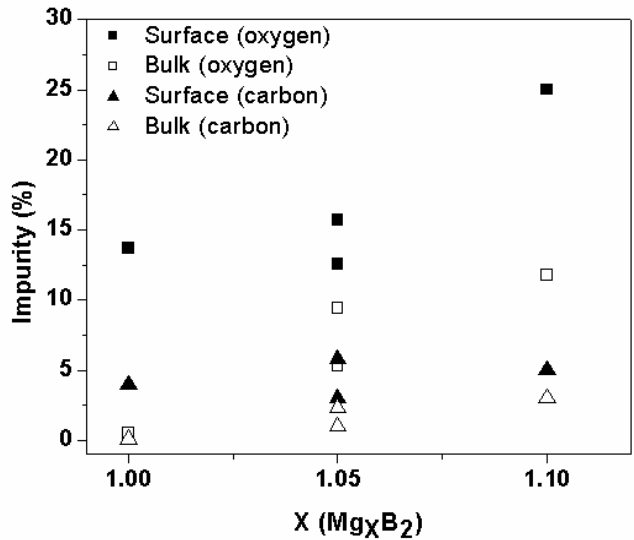

**Figure 6.** Effect of film stoichiometry on oxygen and carbon incorporation after annealing in air at 200 °C for 3 hours.

*3.4 Effect of oxygen on superconducting properties*

Our study on the effect of oxygen incorporation on the superconducting properties of $MgB_2$ films [10] indicates a nominal increase in $\Delta\rho$ with an increase in oxygen concentration. $\Delta\rho$, as pointed out by Rowell [3], is a measure of the inter-grain connectivity in $MgB_2$ samples. An increase in $\Delta\rho$ with oxygen concentration reflects a decrease in the connectivity, presumably as a result of significant oxidation at the grain boundaries. This study [10] also reports a decrease in the $T_c$ with an increase in the oxygen concentration. Oxygen could be incorporated as nm-sized MgO particles in $MgB_2$ grains or be substituted on the boron site in $MgB_2$. It is not clear as to which mechanism is responsible for the $T_c$ suppression. The variations of $T_c$ and $\Delta\rho$ with oxygen concentration indicate the possibility of oxygen incorporation both within and between the grains.

The residual resistivity ($\rho_{40K}$) can be corrected to obtain intra-grain resistivity values using the Rowell analysis [3]. This procedure eliminates any differences in the electrical properties resulting from variations in the inter-grain connectivity. The monotonic increase of corrected residual resistivity ($\rho_{40K,corrected}$) with oxygen concentration [10] is presumably due to increased carrier scattering by MgO particles in $MgB_2$ grains or by oxygen substituted at the boron site.

Our results can be understood as follows: $MgB_2$ has relatively high oxygen solubility at the synthesis temperature of 300 °C [15]. When the temperature is decreased during the cooling process, the solubility of oxygen decreases, forcing the oxygen out of the $MgB_2$ lattice to form Mg(B,O) precipitates. Some oxygen, however, is trapped at interstitial sites. Interstitial oxygen atoms are very mobile, with an activation energy of 1.3 eV [17]. During post-growth annealing of Mg rich samples, diffusing interstitial oxygen combines with Mg in $MgB_2$ and forms MgO precipitates. These incoherent precipitates can act as effective scattering centers.

## 4. Conclusion

It is clear from this study that not only can we control the amount of oxygen incorporated into $MgB_2$, we can also optimize the post growth treatments to control the distribution of oxygen in the matrix. Understanding the mechanism responsible for oxygen incorporation in $MgB_2$ could potentially be used to optimize the scattering and flux pinning for high field applications. Oxygen incorporation in $MgB_2$ films with varying stoichiometry and ambient atmosphere, presented here, is a first step towards that goal.

## Acknowledgements

The authors gratefully acknowledge X.X. Xi (PSU) and B. Moeckly (STI) for providing some of the films used in this study. We also acknowledge use of facilities in the Center for Solid State Science at ASU. This work was supported by NSF Grant No. DMR-0514592 and ONR Contract no. N00014-05-1-0105.

## References

[1] Nagamatsu J, Nakagawa N, Muranaka T, Zenitani Y and Akimitsu J 2001 *Nature* **410** 63-4

[2] Braccini V, Gurevich A, Giencke J E, Jewell M C, Eom C B, Larbalestier D C, Pogrebnyakov A, Cui Y, Liu B T, Hu Y F, Redwing J M, Li Q, Xi X X, Singh R K, Gandikota R, Kim J, Wilkens B, Newman N, Rowell J, Moeckly B, Ferrando V, Tarantini C, Marré D, Putti M, Ferdeghini C, Vaglio R and Haanappel E 2005 *Phys. Rev. B* **71** 012504

[3] Rowell J M 2003 *Supercond. Sci. Technol.* **16** R17-27

[4] Canfield P C, Finnemore D K, Bud'ko S L, Ostenson J E, Lapertot G, Cunningham C E and Petrovic C 2001 *Phys. Rev. Lett.* **86** 2423-6

[5] Zeng X, Pogrebnyakov A V, Kotcharov A, Jones J E, Xi X X, Lysczek E M, Redwing J M, Xu S, Li Q, Lettieri J, Schlom D G, Tian W, Pan X and Liu Z K 2002 *Nature Materials* **1** 35-8

[6] Klie R F, Idrobo J C, Browning N D, Regan K A, Rogado N S and Cava R J 2001 *Appl. Phys. Lett.* **79** 1837-39

[7] Klie R F, Idrobo J C, Browning N D, Serquis A, Zhu Y T, Liao X Z and Mueller F M 2002 *Appl. Phys. Lett.* **80** 3970-2

[8] Gandikota R, Singh R K, Kim J, Wilkens B, Newman N, Rowell J M, Pogrebnyakov A V, Xi X X, Redwing J M, Xu S Y, Li Q and Moeckly B H 2005 *Appl. Phys. Lett.* **87** 072507

[9] Zhu Y, Wu L, Volkov V, Li Q, Gu G, Moodenbaugh A R, Malac M, Suenaga M and Tranquada J 2001 *Physica C* **356** 239-53

[10] Gandikota R, Singh R K, Shen Y, Newman N and Rowell J M  to be submitted

[11] Xue Y Y, Meng R L, Lorenz B, Meen J K, Sun Y Y and Chu C W 2002 *Physica C* **377** 7-14

[12] Kim J, Singh R K, Newman N and Rowell J M 2003 *IEEE Trans. Appl. Supercond.* **13** 3238-41

[13] Kim J, Singh R K, Rowell J M, Newman N, Gu L and Smith D J 2004 *J. Cryst. Growth* **270** 107-12

[14] Moeckly B H and Ruby W S 2006 *Supercond. Sci. Technol.* **19** L21-4

[15] Liao X Z, Serquis A C, Zhu Y T, Huang J Y, Peterson D E and Mueller F M 2002 *Appl. Phys. Lett.* **80** 4398-400

[16] Gandikota R, Singh R K, Kim J, Wilkens B, Newman N, Rowell J M, Pogrebnyakov A V, Xi X X, Redwing J M, Xu S Y and Li Q 2005 *Appl. Phys. Lett.* **86** 012508

[17] Yan Y and Al-Jassim M M 2003 *Phys. Rev B* **67** 212503